\documentclass[nofootinbib,twocolumn,showpacs,aps,prd]{revtex4}

\usepackage{graphicx}
\usepackage{dcolumn}
\usepackage{amsmath}
\usepackage{epsfig}
\usepackage[normalem]{ulem}
\usepackage{epsfig,hhline,amsmath,feynmp}
\usepackage{lscape,dcolumn}
\input pubboard/babarsym
\newcommand{\lumi}    {\ensuremath{384\invfb}\xspace}
\def\llp {\ensuremath{\ellp\ell'^\mp}\xspace}
\def\Lc  {\ensuremath{\Lambda_c^+}\xspace}

\newcommand{\BABARPubYear}    {11}
\newcommand{\BABARPubNumber}  {008}

\newcommand{\SLACPubNumber} {14482}

\def\figurebox#1#2#3{%
    \def\arg{#3}%
    \ifx\arg\empty
    {\hfill\vbox{\hsize#2\hrule\hbox to #2{\vrule\hfill\vbox to #1{\hsize#2\vfill}\vrule}\hrule}\hfill}%
    \else
    {\hfill\epsfbox{#3}\hfill}%
    \fi}

\def\optbar#1{\shortstack{{\tiny (\rule[.4ex]{0.8em}{.1mm})} \\ [-.7ex] $#1$}}

\def\pOrpbar{\kern 0.18em\optbar{\kern -0.18em p}{}\xspace}

\def\pcm {\ensuremath{p^*}\xspace}
\def\Lc  {\ensuremath{\Lambda_c^+}\xspace}
\def\Dsp {\ensuremath{D^+_{(s)}}\xspace}

\def\llp {\ensuremath{\ell^\mp\ell^{(\prime)+}}\xspace}
\def\llpm {\ensuremath{\ell^+\ell^{(\prime)-}}\xspace}
\def\llpp {\ensuremath{\ell^+\ell^{(\prime)+}}\xspace}
\def\DsTopiphi  {\ensuremath{\Ds\to\pip\phi_{KK}}\xspace}
\def\DcTopiphi  {\ensuremath{\Dp\to\pip\phi_{KK}}\xspace}
\def\DspTopiphi  {\ensuremath{\Dsp\to\pip\phi_{KK}}\xspace}
\def\LctopKpi    {\ensuremath{\Lc\to p\Km\pip}\xspace}

\def\DtoPiee {\ensuremath{\Dp\to\pip\epem}\xspace}
\def\DtoPimm {\ensuremath{\Dp\to\pip\mumu}\xspace}

\def\DstoPimm {\ensuremath{\Ds\to\pip\mumu}\xspace}

\def\Xtohll  {\ensuremath{X_c^+\to h^\pm\llp}\xspace}

\newcolumntype{d}[0]{D{.}{.}{-1}}
\newcolumntype{p}[0]{D{p}{\pm}{-1}}

\begin{document}

\preprint{\babar-PUB-\BABARPubYear/\BABARPubNumber} 
\preprint{SLAC-PUB-\SLACPubNumber} 

\begin{flushleft}
\babar-PUB-\BABARPubYear/\BABARPubNumber\\
SLAC-PUB-\SLACPubNumber\\
\end{flushleft}

\title{
{\large \bf
Searches for Rare or Forbidden Semileptonic Charm Decays}
}

%
\author{J.~P.~Lees}
\author{V.~Poireau}
\author{V.~Tisserand}
\affiliation{Laboratoire d'Annecy-le-Vieux de Physique des Particules (LAPP), Universit\'e de Savoie, CNRS/IN2P3,  F-74941 Annecy-Le-Vieux, France}
\author{J.~Garra~Tico}
\author{E.~Grauges}
\affiliation{Universitat de Barcelona, Facultat de Fisica, Departament ECM, E-08028 Barcelona, Spain }
\author{M.~Martinelli$^{ab}$}
\author{D.~A.~Milanes$^{a}$}
\author{A.~Palano$^{ab}$ }
\author{M.~Pappagallo$^{ab}$ }
\affiliation{INFN Sezione di Bari$^{a}$; Dipartimento di Fisica, Universit\`a di Bari$^{b}$, I-70126 Bari, Italy }
\author{G.~Eigen}
\author{B.~Stugu}
\author{L.~Sun}
\affiliation{University of Bergen, Institute of Physics, N-5007 Bergen, Norway }
% ...
\collaboration{The \babar\ Collaboration}
\noaffiliation

\date{\today}

\begin{abstract}
We present searches for rare or forbidden charm decays of the form \Xtohll,
where $X_c^+$ is a charm hadron (\Dp, \Ds, or \Lc), 
$h^\pm$ is a pion, kaon, or proton, and
$\ell^{(\prime)\pm}$ is an electron or muon. 
The analysis is based on \lumi of \epem annihilation data collected at
or close to the \Y4S resonance with the \babar\ detector at
the SLAC National Accelerator Laboratory.  No significant signal is
observed for any of the 35 decay modes that are investigated.  We
establish 90\% confidence-level upper limits on the branching
fractions between $1 \times 10^{-6}$ and $44 \times 10^{-6}$ depending
on the channel.  In most cases, these results represent either the
first limits or significant improvements on existing limits for the decay modes studied.
\end{abstract}


\maketitle

\section{Introduction}

\label{sec:introduction}

We present searches for charm hadron decays that are either forbidden
or heavily suppressed in the Standard Model (SM) of particle physics.
The decays are of the form~\footnote{The inclusion of charge-conjugate decay modes
is implied throughout this paper.} \Xtohll, where $X_c^+$
is a charm hadron (\Dp, \Ds, or \Lc),
and $\ell^{(\prime)\pm}$ is an electron or muon. For \Dp and \Ds modes, $h^\pm$  can be a pion or kaon, while for \Lc modes it is a proton.
 Decay modes with
oppositely charged leptons of the same lepton flavor are examples of
flavor-changing neutral current (FCNC) processes, which are expected
to be very rare because they cannot occur at tree level in the SM.
Decay modes with two oppositely charged leptons of different
flavor correspond to lepton-flavor violating (LFV) decays and are
essentially forbidden in the SM because they can occur only through
lepton mixing.  Decay modes with two leptons of the same charge
are lepton-number violating (LNV) decays and are forbidden in the SM.
Hence, decays of the form \Xtohll\ provide sensitive tools to
investigate physics beyond the SM. The
most stringent existing upper
limits~\cite{Rubin:2010cq,Dzero,focus,Aitala:1999db,Kodama:1995ia} on
the branching fractions for \Xtohll\ decays range from 1 to
$700\times10^{-6}$ and do not exist for most of the $\Lc$ decays.

FCNC processes have been studied extensively for $K$ and $B$ mesons, in
$\Kz-\Kzb$ and $\Bz-\Bzb$ mixing, and in rare FCNC decays
such as $s\to d\ellell$, $b\to s\gamma$, and $b\to s\ellell$ decays.
The results
agree with expectations within the framework of the SM \cite{Antonelli:2009ws}.
There are ongoing efforts to improve the measurements
and the theoretical predictions, and to measure new effects, such as
\CP violation, in FCNC processes.

\begin{figure}[t]
\begin{fmffile}{ctoullWbox} 
  \fmfframe(20,20)(10,1){  
   \begin{fmfgraph*}(180,100)
    \fmfstraight
    \fmfleft{i0,i1,i2,i3,i4,i5,i6,i7}
    \fmfright{o0,o1,od,od2,od3,o2,od4,o3}
    \fmftop{t1,t2,t3,t4}
    \fmflabel{$c$}{i1}
    \fmflabel{$u$}{o1}
    \fmflabel{$\ell$}{o2}
    \fmflabel{$\ell$}{o3}
    \fmf{fermion}{i1,v1,v2,o1}
    \fmffreeze
    \fmf{boson,label=$W$}{v1,v3}
    \fmf{boson,label=$W$}{v2,v4}
    \fmf{phantom}{v3,m3,t2}
    \fmf{phantom}{v4,m4,t3}
    \fmffreeze
    \fmf{fermion}{o2,v4,v3,o3}
   \end{fmfgraph*} 
  }
\end{fmffile}
\begin{fmffile}{ctoullPeng}  
  \fmfframe(10,1)(1,4){ 
   \begin{fmfgraph*}(180,100) 
    \fmfstraight
    \fmfleft{i0,i1,i2,i5,i6,i7} 
    \fmfright{o0,o1,od,o2,o3,od4} 
    \fmftop{t1,t2,t3,t4}
    \fmflabel{$c$}{i1} 
    \fmflabel{$u$}{o1}
    \fmflabel{$\ell$}{o2}
    \fmflabel{$\ell$}{o3}
    \fmf{fermion}{i1,v1}
    \fmf{plain}{v1,v3}
    \fmf{fermion}{v3,v4}
    \fmf{plain}{v4,v2}
    \fmf{fermion}{v2,o1} 
    \fmf{boson,right,label=$W$}{v1,v2}
    \fmffreeze
    \fmf{boson,label=$\gamma/Z^0$,label.side=left}{v4,v5}
    \fmf{fermion}{o3,v5,o2}
    \fmf{phantom}{t3,v5}
   \end{fmfgraph*} 
  }
\end{fmffile}
\caption{\label{fig:Feynman} Standard model short-distance
contributions to the $c\to u\ellell$ transition.}
\end{figure}
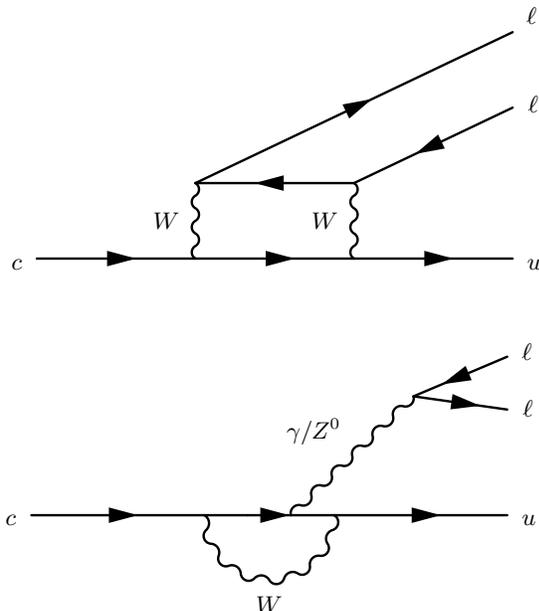

The recent observation of $\Dz-\Dzb$ mixing~\cite{Aubert:2007wf,Staric:2007dt,Aaltonen:2007uc} has increased interest in FCNC
processes in the charm sector. Of particular interest is the source
of  $\Dz-\Dzb$ mixing. If the mixing is due to physics beyond the SM, it
could also give rise to measurable effects in FCNC charm decays.  In
the SM, the FCNC decays $X_c^+ \to h^+ \ell^+\ell^-$ are expected to
be heavily suppressed due to cancellations of amplitudes through the
Glashow-Iliopoulos-Maiani (GIM) mechanism~\cite{Glashow:1970gm}.  For example, the $c\to
u\ellell$ transitions illustrated in Fig.~\ref{fig:Feynman} yield
branching fractions for $D\to X_u\ellell$ of ${\cal{O}}(10^{-8})$~\cite{Burdman,Fajfer}. These decays are masked by the presence of
long-distance contributions from intermediate vector resonances such
as $D\to X_u V,$ $V\to\ellell$, which are predicted to have
branching fractions of ${\cal{O}}(10^{-6})$~\cite{Burdman,Fajfer}. 
The effect of these resonances can be separated from those due
to short-range processes by applying selection criteria on the 
invariant mass of
the $\ellell$ pair. In radiative charm decays, $c\to u\gamma$, uncertainties in calculating 
the long-distance terms make it impossible to study the underlying
short-distance physics~\cite{BurdmanRadiative}.

The impact of several extensions of the SM on $D\to X_u\ellell$ decay rates 
has been estimated~\cite{Burdman,LittleHiggs,Paul:2011ar}; the largest
effect not already ruled out is expected in certain R-parity violating supersymmetric
models. Depending on the size of the R-parity violating couplings,
branching fractions of up to ${\cal{O}}(10^{-5})$ for different $D\to
X_u\ellell$ decays are possible. 

The only long-distance amplitudes relevant at the current
experimental sensitivity are from $\Dp\to\pip\phi$ and
$\Ds\to\pip\phi$ decays. The product of branching fractions,
${\cal B}(\Dsp\to\pip\phi){\cal B}(\phi\to\ellell)$, is
$\approx 2\times10^{-6}$ for $\Dp$ and $\approx 1\times10^{-5}$ for $\Ds$~\cite{PDG}. 
In this analysis, we measure the total decay rate of $\Dsp\to\pip\ellell$
excluding the \ellell mass region around the $\phi$ resonance.

\section{Overview}

\label{sec:overview}

The searches use data collected with the \babar{} detector as described
in Section~\ref{sec:dataset}. Candidate decays are formed from two
tracks identified as leptons ($\ell\ell^{(\prime)}$) and one track identified as a
hadron ($h$). Background events arise primarily from semileptonic $B$ and charm decays and from radiative
QED events with converted photons. After applying basic kinematic
selection criteria, we discriminate further between signal and background
with a decay-mode-dependent
likelihood ratio calculated from the measured momentum and flight length of the
charm hadron candidate and the total energy detected in the event. The invariant mass
distributions of the selected $h\ell\ell^{(\prime)}$ candidates are fit to extract the
number of signal events, the number of combinatorial background events,
 and the number of background events due to nonleptonic charm decays 
 with hadrons in the final state misidentified as leptons.  
 Upper limits are determined for the decay rate to each final state with respect to
 the known decay rate for a hadronic three-body decay with similar kinematics so that many 
 systematic uncertainties cancel in the ratio.

\section{The \babar\ detector and dataset}

\label{sec:dataset}
The \babar{} detector was operated at the \pep2\ asymmetric-energy storage
rings at the SLAC National Accelerator Laboratory. The data sample
comprises an integrated luminosity of 347\invfb collected from \epem
collisions at the \FourS resonance and 37\invfb collected 40\mev below
the \FourS resonance.

The \babar\ detector is described in detail
elsewhere~\cite{ref:babar}.  The momenta of charged particles are
measured with a combination of a five-layer silicon vertex tracker
(SVT) and a 40-layer drift chamber (DCH), both in a 1.5~T magnetic field
produced by a solenoid.  The resolution of the transverse momentum (\pt) is measured to be
$\sigma(\pt)/\pt=0.0013(\pt/\gevc)\oplus 0.0045$. A detector of
internally reflected Cherenkov radiation (DIRC) is used for charged
particle identification.  Pions, kaons, and protons are identified
with likelihood ratios calculated from $dE/dx$ measurements in the SVT
and DCH, and from the observed pattern of Cherenkov light in the DIRC.  For
the selection criteria used in this analysis, hadron identification
efficiencies are approximately 98\%, 87\%, and 82\% for pions, kaons,
and protons, respectively.
A finely segmented CsI(Tl) electromagnetic calorimeter (EMC) is used
to detect and measure the energies of photons and neutral hadrons, and
to identify electrons. For electrons, energy lost due bremsstrahlung
is recovered from deposits in the EMC.

The EMC information, Cherenkov angle, and $dE/dx$ measurements are combined to define a likelihood ratio used to select electrons with a 
selection efficiency above $90\%$ for electrons with a laboratory momentum
above 0.5\gevc and with less than $0.1\%$ probability to misidentify a
hadron as an electron. The instrumented flux return (IFR) contains
resistive plate chambers and limited streamer tubes for muon and
long-lived neutral-hadron identification. Variables
characterizing track measurements in the IFR and the energy deposition
in the EMC are combined in a neural network to select muon
candidates. The muon identification efficiency is about 60\% for muons
with a laboratory momentum above 1.5\gevc, but decreases rapidly for lower
momenta.  The probability to misidentify a pion as a muon is about
1.5\% for most of the relevant momentum range.

Event simulations are performed using the EvtGen~\cite{evtgen} Monte
Carlo (MC) generator with a full detector simulation based on
GEANT4~\cite{geant4}.  Signal and \LctopKpi MC events are generated
with a three-body phase-space distribution, while $\Dsp\to\pip\phi$ MC
events are generated with a Breit-Wigner resonance shape for the
$\phi$ decay and an angular distribution appropriate for the P-wave $\phi$
decay. All signal events are simulated as \ccbar continuum
events.  The distributions of the magnitudes of the charm-hadron momenta  measured in the \epem center-of-mass frame (\pcm) are found to differ between simulated \ccbar continuum events and data. 
In particular, the
mean \pcm for \Ds mesons is lower in MC, while the mean \pcm for \Lc
baryons is higher.  To correct for this, simulated \ccbar events are
weighted to yield the same \pcm distributions as those observed for
large samples of reconstructed charm decays in data. The event weights
lie between 0.6 and 1.8. Simulated events are also weighted to match
the particle identification probabilities measured in control samples
in data.  Samples of simulated \ccbar and $q \overline{q}$ ($q=u,d,s$)
continuum events and \BB events, corresponding to 1.4 to 5 times the
recorded data sample, are used to study background contributions.

\section{Analysis procedure}

\label{sec:analysis}

\subsection{Initial Signal Selection}

\label{sec:initialselection}
Charm hadron candidates are formed from one track identified
as either a pion, kaon, or proton ($h$) and two tracks, each of which is
identified as an electron or a muon ($\ell\ell^{(\prime)}$). 
The total charge of the three tracks is required to be $\pm1$. 
For three-track combinations with a pion or kaon track, the $h\ell\ell^{(\prime)}$ 
invariant mass is required to lie between 1.7
and 2.1\gevcc; for combinations with a proton, the invariant mass is required to
lie between 2.2 and 2.4\gevcc.  

The combinatorial background at low \pcm is very large and we therefore select
charm hadron candidates with \pcm greater than 2.5\gevc.
Because the \pcm of charm hadrons produced in \B decays is kinematically
limited to be less than about 2.2 \gevc, the signal candidates with
\pcm greater than 2.5\gevc are dominated by hadrons from continuum
production.  The main backgrounds remaining after this selection are
QED events and semileptonic $B$ and charm decays, particularly events
with two semileptonic decays.

The QED events are mainly radiative Bhabha, initial-state radiation,
and two-photon events, which are all rich in electrons.  These events
are easily identified by their low multiplicity and/or highly jet-like
topology. We strongly suppress this background by requiring at least five
tracks in the event and that the hadron candidate  
be inconsistent with the electron hypothesis.  

We suppress the background from semileptonic $B$ and charm decays by
requiring the two leptons to be consistent with a common origin.  This is
achieved by requiring that the probability of the $\chi^2$ for the vertex fit 
be greater than $0.001$ and that the distance of closest approach
between the two lepton candidates be less than $250\mum$.

For low \epem invariant mass there is a significant background contribution
from photon conversions and $\piz$ decays to $\epem\gamma$.
These are both removed by requiring $m(\epem)>200\mevcc$.

For the $\Dsp\to\pip\ellell$ decay modes, we exclude events with
$0.95<m(\epem)<1.05\gevcc$ and $0.99<m(\mumu)<1.05\gevcc$ to reject
decays through the $\phi$ resonance. The excluded regions for the two
decay modes differ due to the larger radiative tail in the
$\pip\epem$ decay mode. In order to perform cross-checks, we also
select candidates for  $\Dsp\to\pi\phi$, $\phi\to\ellell$ decays by 
requiring $0.995\gevcc<m(\epem)<1.030\gevcc$ or
$1.005\gevcc<m(\mumu)<1.030\gevcc$.

\subsection{Likelihood Selection}

\label{sec:likelihoodselection}
After the initial event selection, significant combinatorial
background contributions remain from semileptonic \B decays and other
sources.  These background sources are studied with samples of
candidates in MC and in sidebands of the $h\ell\ell^{(\prime)}$
invariant mass distribution in data. The sidebands are defined to be
0.1~\gevcc wide regions just below and above the signal regions defined above.
 The final candidate selection is
performed by forming a likelihood ratio $R_{\cal{L}}$ and requiring
the ratio to be greater than a minimum value $R^{\rm{min}}_{\cal{L}}$.
The likelihood ratio is defined as
\begin{equation}
R_{\cal{L}}(\vec{x})=\frac{\prod_i {\cal P}_{s,i}(x_i)}{\prod_i {\cal P}_{s,i}(x_i)+\prod_i {\cal P}_{b,i}(x_i)},
{\label{eqn:likelihoodratio}} 
\end{equation}
where $x_i$ is the $i$th discriminating variable, and
${\cal P}_{s,i}$ and ${\cal P}_{b,i}$ are the signal and
background probability
density functions (PDF) for the variable $x_i$. Correlations between
variables are ignored in the likelihood ratio as they are found to be small.
The likelihood ratio peaks
near 1 for signal and near 0 for background. The signal PDFs are
obtained by fitting the distribution of $x_i$ for
simulated signal events, while the background PDFs
are obtained from fits to the distribution of $x_i$ for candidates in 
the $h\ell\ell^{(\prime)}$ invariant-mass sidebands in data. The PDFs are defined by combinations
of polynomial, Gaussian, and exponential functions found empirically to provide good descriptions of the signal MC and data sidebands.
 The parameter values in the PDFs are
common across signal modes in which data or MC studies show the
distributions to be consistent.

\begin{figure}[htb]
\centerline{\includegraphics[width=9cm]{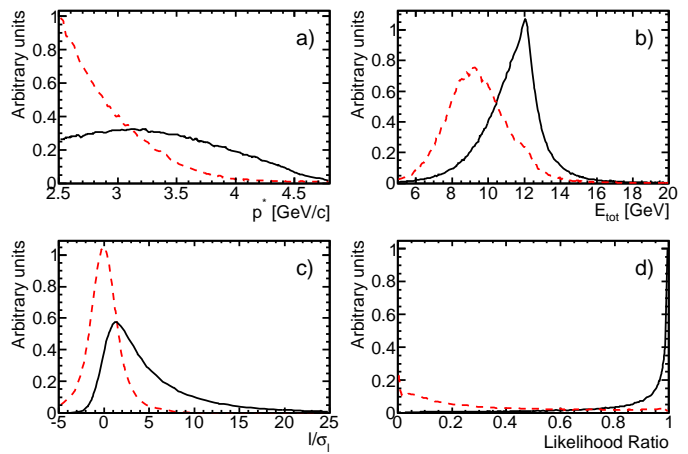}}
\caption{\label{fig:LikelihoodExample} 
Distributions of the three discriminating variables
and the likelihood ratio, for  \DtoPiee candidates in signal MC (black solid curve) and 
in the $h\ell\ell^{(\prime)}$ invariant-mass sidebands from data (red dashed curve): 
a) Center-of-mass momentum \pcm of the charm-hadron candidate;
b) Total energy in the event;  
c) Flight length significance; 
d) Likelihood ratio (defined in Eq.~\ref{eqn:likelihoodratio}) calculated with the three discriminating variables.
The signal and background distributions are normalized to the same area.}
\end{figure}

The following three discriminating variables are used in the likelihood ratio.
\begin{itemize}
\item Charm hadron candidate \pcm.\\ 
      The calculated \pcm for $h\ell\ell^{(\prime)}$ candidates in which a lepton candidate is from a semileptonic $B$ decay can be larger than 2.2\gevc, but the distribution
      falls rapidly with increasing \pcm.
\item Total reconstructed energy in the event.\\ 
      Since neutrinos from semileptonic $B$ decays are not directly
      detected, the total observable energy in these
      events is less than the sum of the beam energies (12\gev).  We calculate
      the total energy for each event as the sum of the energies of 
      all reconstructed tracks (assuming each track to be a charged pion) and
      neutral EMC clusters.
 \item Flight length significance.\\
      The flight length significance is the ratio of the signed flight length
      to its uncertainty, where the signed flight length is the scalar product
      between the direction of the charm-hadron candidate and the position 
      vector that points from the beam spot to the charm-hadron
      decay vertex. 
      This variable has the most discrimination power for the long-lived
      \Dp meson and the least for the shorter-lived \Lc. 
      This variable is effective for suppressing non-$B$ combinatorial background. 
\end{itemize}
Distributions of the three discriminating variables and of the likelihood ratio
are shown in Fig.~\ref{fig:LikelihoodExample}  for \DtoPiee candidates in 
$h\ell\ell^{(\prime)}$ invariant-mass sidebands (background) and in signal MC.

The minimum likelihood ratio value $R^{\rm{min}}_{\cal{L}}$ is chosen 
independently for each mode to provide the
lowest expected upper limit on the branching ratio, as calculated from
the simulated signal efficiency and the expected number of background events; the latter is estimated
from the $h\ell\ell^{(\prime)}$ invariant-mass sidebands in data. 
In cases in which more than one candidate
from the same event passes all selection criteria, including the
likelihood-ratio requirement, the candidate with the highest \pcm is
retained. The final signal selection efficiency lies between 0.5\% and 7\%,
depending on the signal mode.

\subsection{Fit Procedure}

\label{sec:fitprocedure}

Extended, unbinned, maximum-likelihood fits are applied to the
invariant-mass distributions for the $h^\pm\llp$ candidates.  The PDF
we use for signal events is the so-called Crystal Ball
function~\cite{CBPaper}, which has an asymmetric component to
describe the radiative tail in the mass distribution:
\begin{eqnarray*}
  \lefteqn{P_{CB}(m;\mu,\sigma,\alpha,n)=}\\
& & \left\{\begin{array}{ll}
    e^{-\frac{1}{2}(\frac{m-\mu}{\sigma})^2} & \mbox{if }m\ge \mu-\alpha\sigma,\\
    e^{-\frac{1}{2}\alpha^2}\left(\frac{n\sigma}{n\sigma-\alpha^2\sigma-\alpha(m-\mu)}\right)^n & \mbox{if } m< \mu-\alpha\sigma.
  \end{array}\right.
\end{eqnarray*}
The four parameters $\mu,\sigma,\alpha$ and $n$ are determined from
fits to signal MC candidates and are fixed to these values in the fits to data, with only
the overall normalization as a free parameter.  The width of
the Gaussian component ($\sigma$) is found to lie between 5 and
10\mevcc, depending on the decay mode.

The invariant mass distributions of the combinatoric background events
for the signal modes are described by first-order polynomials.  The
background slope is a free parameter in all fits.

An additional background arises from nonleptonic charm decays in which two
hadrons are misidentified as leptons.
This background component is almost
negligible in the signal modes with electrons and is therefore only
included in decay modes with two muons and for $\Dsp\to\Km\mup\ep$ where 
there is a large $\Dp\to\Kp\pip\pip$ background. The shape of this background
is obtained from MC samples of major hadronic three-body charm decays. Each
event is weighted according to the probability of misidentifying a
pion as a lepton. The misidentification probability is measured from
data as a function of momentum and angle with samples of $\Dz\to\Km\pip$ candidates. The misidentified
nonleptonic charm decays are reconstructed at slightly lower $h^+\mumu$ ($\Km\mup\ep$)
mass than the signal events. The peak mass is shifted by about
15\mevcc, which is sufficient separation from the signal peak 
for this background yield to be determined 
by the likelihood fit to data, without reliance on MC predictions of this yield.

\section{Branching Ratio Normalization}

\label{sec:normalization}

The measured signal yields are converted into branching ratios by
normalizing them to the yields of known charm decays.  We choose
normalization modes with kinematics similar to the kinematics of the signal decays
so that most systematic effects not related to particle
identification cancel in the branching ratio. 
For the \Dp and \Ds mesons,
we use decays to $\pip\phi$ as normalization modes, where the $\phi$ decays to
$\KpKm$. For the \Lc, we use $\Lc\to\proton\Km\pip$ as the normalization mode. 
The measured branching fractions for these modes are listed in
Table~\ref{tab:normBF}. We use the abbreviation \DspTopiphi to denote
$\Dsp\to\pip\phi, \phi\to\KpKm$ decays. 

\begin{table}
\caption{\label{tab:normBF} Branching fractions for
the charm decays used for normalization~\cite{PDG}.}
\begin{center}
\begin{tabular*}{0.48\textwidth}{@{\extracolsep{\fill}}lp} \hline \hline
Decay mode                       &  \multicolumn{1}{r}{Branching Fraction}  \\ \hline
\rule{0cm}{2.3ex}$\DcTopiphi$  & (2.72 p 0.13)\times10^{-3} \\ 
$\DsTopiphi$ & (2.32  p 0.14)\times10^{-2} \\ 
$\Lc\to\proton\Km\pip$           & (5.0  p 1.3)\times10^{-2}  \\ \hline \hline
\end{tabular*}
\end{center}
\end{table}

\subsection{Event Selection}

The same selection criteria are applied for the normalization modes as
the signal modes, except for the lepton identification and likelihood ratio
requirements. 
Instead, each daughter candidate in the \DspTopiphi
and \LctopKpi decay modes is required to be consistent with 
the kaon, pion, or proton hypothesis, as appropriate.
For the \DspTopiphi decay modes, we further require the
invariant mass of the kaon pair to lie within 15\mevcc of the
world-average $\phi$ mass~\cite{PDG}. 

\subsection{Fit Results}

\begin{figure}[htb]
\centerline{\includegraphics[width=9cm]{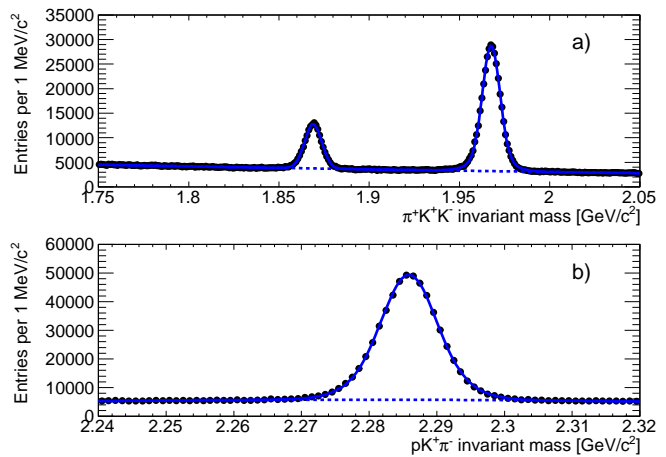}}
\caption{\label{fig:RefFit} Invariant mass distribution for a)
  $\pip\phi_{\Kp\Km}$ and b) $pK\pi$ candidates. The solid lines
  are the results of fits to double-Gaussian distributions for signals and a
  second-order polynomial for the background (dashed line).}
\end{figure}

For the normalization modes,  radiative effects are
negligible and we use the sum of two Gaussian distributions with a
common mean to describe each of the \Dp, \Ds, and \Lc  signal invariant-mass distributions. 
All signal parameters are free in the fits to the normalization modes in data. 
The combinatorial background is described by a second-order polynomial.
The invariant-mass distributions for the normalization decay modes and
the corresponding fit results are shown in Fig.~\ref{fig:RefFit}. The fitted
signal yields are listed in Table~\ref{tab:referenceFits}, where we
also list the efficiencies estimated from signal MC. 

The \DspTopiphi samples include a small component of non-resonant or S-wave
\Dsp{} decays, while the branching fractions in Table~\ref{tab:normBF}
are extracted from decay amplitude  analyses of Dalitz  plot distributions for $\Dsp\to\pip\KpKm$ decays and
therefore correspond only to the resonant component. This component
is estimated in our MC and data samples by projecting out the P-wave component
by weighting each event with a factor that includes the reciprocal of its reconstruction
efficiency and a normalized L=2 Legendre polynomial function of the $\KpKm$ helicity angle.
The fractions of \Dp and \Ds decays found to proceed through a P-wave
are 94\% and 93\%, respectively, and are used to correct the fit yields.

\begin{table}
\caption{\label{tab:referenceFits} Fitted signal yields for the
  normalization modes and signal efficiencies estimated from MC simulations. 
   Only statistical uncertainties are quoted.}
\begin{center}
\begin{tabular*}{0.48\textwidth}{@{\extracolsep{\fill}}lcc} \hline \hline
Decay mode &  $N_{\mathrm{sig}}$ & Efficiency \\ \hline
\rule{0cm}{2.3ex}\DcTopiphi & $106\,800\pm500$     & $(15.44\pm0.07)$\% \\
\DsTopiphi & $338\,900\pm900$     & $(15.29\pm0.07)$\% \\
\LctopKpi  & $488\,700\pm2\,100$    & $(11.99\pm0.04)$\% \\ \hline \hline
\end{tabular*}
\end{center}
\end{table}

\section{Systematic Uncertainties}

\label{sec:systematics}

Most systematic effects are expected to cancel in the branching ratio
since they affect both the signal and normalization modes. 
Therefore, the main systematic uncertainties arise from differences in selection,
acceptance, and decay kinematics.  Table~\ref{tab:Systematics} gives a
summary of all systematic uncertainties related to the branching
ratio calculation. An additional systematic uncertainty is associated with the
estimate of the signal yield. 

\begin{table}
\caption{\label{tab:Systematics} Summary of the multiplicative systematic uncertainties 
on the branching ratio for each signal decay mode (listed in the first column).
In the second column, we give the fractional uncertainty due to the normalization mode.
The next three columns list the fractional uncertainty due to the MC statistical error, 
uncertainties in particle identification efficiencies (PID), and in the likelihood-ratio efficiency (LR).
The total systematic uncertainty, given in the final column, is the sum of the individual errors in quadrature.
}
\begin{center}
\begin{tabular*}{0.48\textwidth}{@{\extracolsep{\fill}}lccccc} \hline \hline
           & Norm. &          &      &      &        \\
Decay mode &       Mode    & MC stat. &  PID &  LR & Total  \\ \hline
\rule{0cm}{2.3ex}
$\Dp\to\pip\epem$           & 2.6\% & 1.8\% & 2.3\% & 9.8\% & 10.6\% \\
$\Dp\to\pip\mumu$           & 2.6\% & 4.6\% & 8.3\% & 6.1\% & 11.6\% \\
$\Dp\to\pip\ep\mun$         & 2.6\% & 3.1\% & 4.4\% & 9.8\% & 11.5\% \\
$\Dp\to\pip\mup\en$         & 2.6\% & 2.2\% & 4.4\% & 7.6\% & 9.4\% \\
$\Ds\to\pip\epem$           & 2.1\% & 1.1\% & 2.3\% & 0.9\% & 3.4\% \\
$\Ds\to\pip\mumu$           & 2.1\% & 2.7\% & 8.3\% & 0.9\% & 9.0\% \\
$\Ds\to\pip\ep\mun$         & 2.1\% & 2.6\% & 4.4\% & 2.9\% & 6.2\% \\
$\Ds\to\pip\mup\en$         & 2.1\% & 4.0\% & 4.4\% & 7.2\% & 9.6\% \\
$\Dp\to\Kp\epem$            & 2.6\% & 1.4\% & 2.8\% & 5.5\% & 6.8\% \\
$\Dp\to\Kp\mumu$            & 2.6\% & 6.4\% & 8.5\% & 4.4\% & 11.8\% \\
$\Dp\to\Kp\ep\mun$          & 2.6\% & 2.9\% & 4.7\% & 5.8\% & 8.4\% \\
$\Dp\to\Kp\mup\en$          & 2.6\% & 3.1\% & 4.7\% & 5.1\% & 8.0\% \\
$\Ds\to\Kp\epem$            & 2.1\% & 1.5\% & 2.8\% & 2.0\% & 4.3\% \\
$\Ds\to\Kp\mumu$            & 2.1\% & 3.3\% & 8.5\% & 0.9\% & 9.4\% \\
$\Ds\to\Kp\ep\mun$          & 2.1\% & 2.2\% & 4.7\% & 2.0\% & 5.9\% \\
$\Ds\to\Kp\mup\en$          & 2.1\% & 2.0\% & 4.7\% & 1.6\% & 5.7\% \\
$\Lc\to\proton\epem$        & 3.4\% & 1.4\% & 2.0\% & 5.4\% & 6.8\% \\
$\Lc\to\proton\mumu$        & 3.4\% & 3.2\% & 8.2\% & 3.4\% & 10.0\% \\
$\Lc\to\proton\ep\mun$      & 3.4\% & 3.4\% & 4.3\% & 5.6\% & 8.5\% \\
$\Lc\to\proton\mup\en$      & 3.4\% & 2.8\% & 4.3\% & 5.1\% & 8.0\% \\ \hline

$\Dp\to\pim\ep\ep$          & 2.6\% & 1.3\% & 2.3\% & 5.7\% & 6.8\% \\
$\Dp\to\pim\mup\mup$        & 2.6\% & 3.5\% & 8.3\% & 5.1\% & 10.7\% \\
$\Dp\to\pim\mup\ep$         & 2.6\% & 1.6\% & 4.4\% & 4.6\% & 7.1\% \\
$\Ds\to\pim\ep\ep$          & 2.1\% & 0.8\% & 2.3\% & 1.2\% & 3.4\% \\
$\Ds\to\pim\mup\mup$        & 2.1\% & 2.2\% & 8.3\% & 1.2\% & 8.9\% \\
$\Ds\to\pim\mup\ep$         & 2.1\% & 2.4\% & 4.4\% & 1.2\% & 5.6\% \\
$\Dp\to\Km\ep\ep$           & 2.6\% & 1.4\% & 2.8\% & 5.7\% & 7.0\% \\
$\Dp\to\Km\mup\mup$         & 2.6\% & 3.2\% & 8.5\% & 2.7\% & 9.8\% \\
$\Dp\to\Km\mup\ep$          & 2.6\% & 2.4\% & 4.7\% & 3.9\% & 7.1\% \\
$\Ds\to\Km\ep\ep$           & 2.1\% & 1.0\% & 2.8\% & 1.4\% & 3.9\% \\
$\Ds\to\Km\mup\mup$         & 2.1\% & 2.8\% & 8.5\% & 0.6\% & 9.2\% \\
$\Ds\to\Km\mup\ep$          & 2.1\% & 1.9\% & 4.7\% & 0.6\% & 5.5\% \\
$\Lc\to\antiproton\ep\ep$   & 3.4\% & 1.0\% & 2.0\% & 5.4\% & 6.8\% \\
$\Lc\to\antiproton\mup\mup$ & 3.4\% & 2.6\% & 8.2\% & 1.7\% & 9.4\% \\
$\Lc\to\antiproton\mup\ep$  & 3.4\% & 1.6\% & 4.3\% & 3.5\% & 6.7\% \\ \hline

\end{tabular*}
\end{center}
\end{table}

Systematic uncertainties related to the signal PDF parameters are
investigated in two ways. First, the PDF parameters for data and MC
are compared in the fits to the normalization modes. Differences can
be due either to general data-MC tracking differences or to
differences between the simulated charm hadron mass and the actual
mass. Second, fits to the invariant mass of $\jpsi\to\ellell$
candidates from inclusive \B decays are compared between data and
MC. The second comparison is sensitive to effects associated with
lepton reconstruction and radiative tails. Based on these studies, the
mean signal mass is varied by up to 2.5\mevcc, depending on the decay
mode. Based on the same studies, the widths of the signal PDFs are
varied by $\pm 5\%$. For the background shape assumption, the signal
fits are repeated using a second-order polynomial as the background
PDF instead of the nominal first-order polynomial. The PDF variation
giving the worst upper limit on the branching ratio is used as an estimate
of the systematic uncertainty.

For the normalization modes, the statistical uncertainties from the
fits, the MC statistical errors, and uncertainties from the signal and
background shapes are all about 1\% or less. The main systematic uncertainty is due to
intermediate resonances in the three-body decays. 
For the \Dsp modes, where we select $\KpKm$ pairs consistent with the $\phi$ mass, this uncertainty is
estimated by varying the \KpKm mass interval in order to alter the purity of the accepted $\phi$ candidates.
For the \Lc mode, the uncertainty is estimated from variations in the efficiency
as a function of the $\proton\Km$ and $\proton\pip$ invariant mass.
For all three normalization modes the estimated uncertainty is approximately 2\%.

The particle identification efficiencies have associated
systematic uncertainties of 0.5\% for each pion, 0.7\% for
each kaon, 0.9\% for each electron, and 4\% for each muon. We do not
evaluate a systematic uncertainty for the protons since, for the $\Lambda_c$ decays, 
both the signal
and normalization modes contain a proton and the uncertainty
therefore cancels.  Uncertainties for particles of the same type are
added linearly, while those for different types of particles are added in
quadrature.

The systematic uncertainty on the efficiency of the likelihood-ratio requirement
is estimated by applying the same likelihood ratio
selection to events in the normalization mode. The efficiency of
this selection for the normalization-mode decays is not expected to be the same
as for signal modes due to different kinematics. However, for the normalization modes, the
efficiency can be measured for both data and for MC simulation; the difference
is used as the systematic uncertainty. The largest variations are found for
decay modes with the most stringent likelihood-ratio requirements.

In the calculation of the signal efficiency we assume that the decays
populate three-body phase-space uniformly. The selection
efficiency has some dependence on where the decay lies in the Dalitz
plot. Ignoring the regions we explicitly remove in the selection, the
efficiency varies by less than 25\% around the average as a function
of the $\llp$ invariant mass, with the lowest efficiency at low
dilepton mass. This model dependence is not included in the systematic
uncertainty.

\section{Results}

\label{sec:results}

\begin{figure*}[tb]
\centerline{\includegraphics[width=18cm]{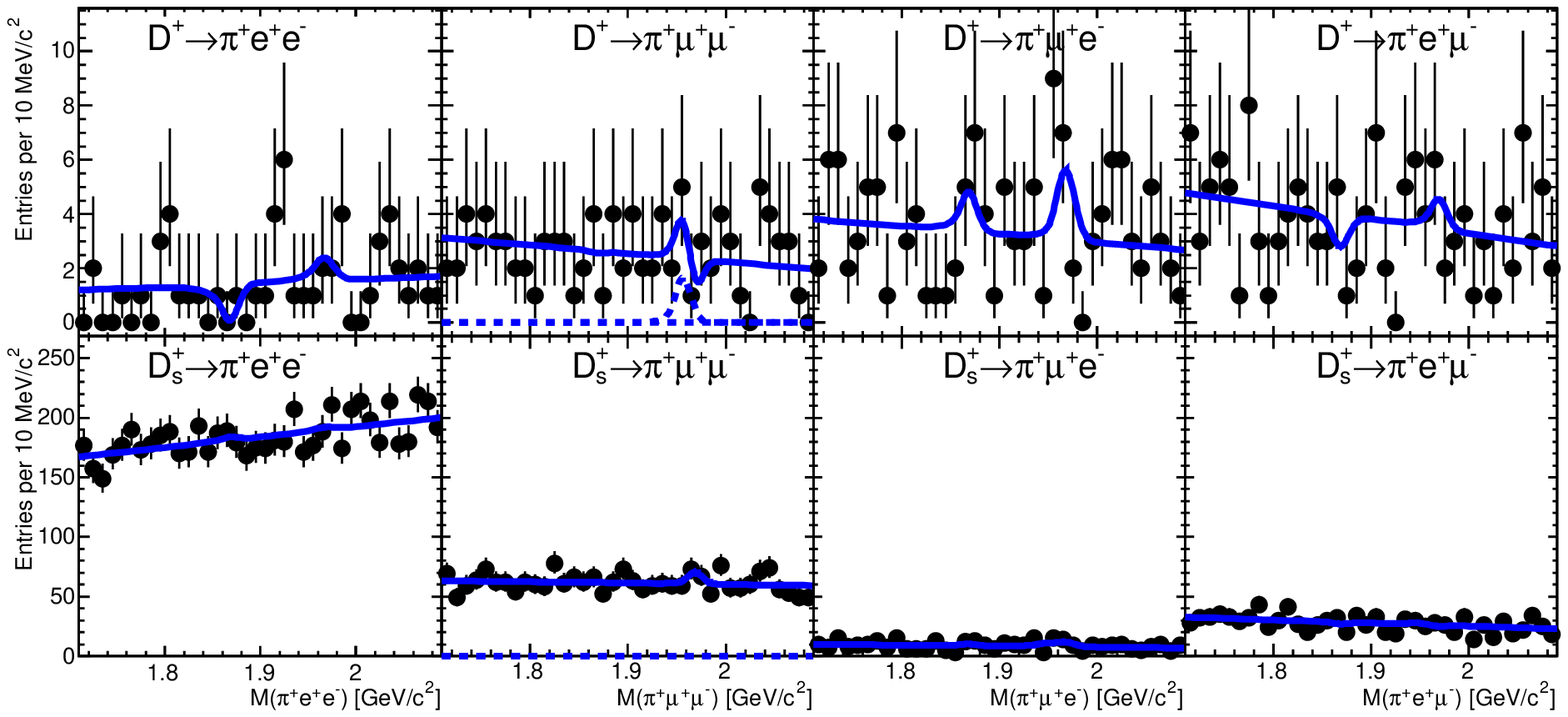}}
\caption{\label{fig:fitDppill} Invariant-mass distributions for
$\Dp\to\pip\llpm$ (top) and $\Ds\to\pip\llpm$ (bottom) candidates.
The solid lines are the results of the fits. 
The background components for the dimuon modes in which muon candidates arise from misidentified hadrons are shown as dashed curves.}
\end{figure*}

\begin{figure*}[tb]
\centerline{\includegraphics[width=18cm]{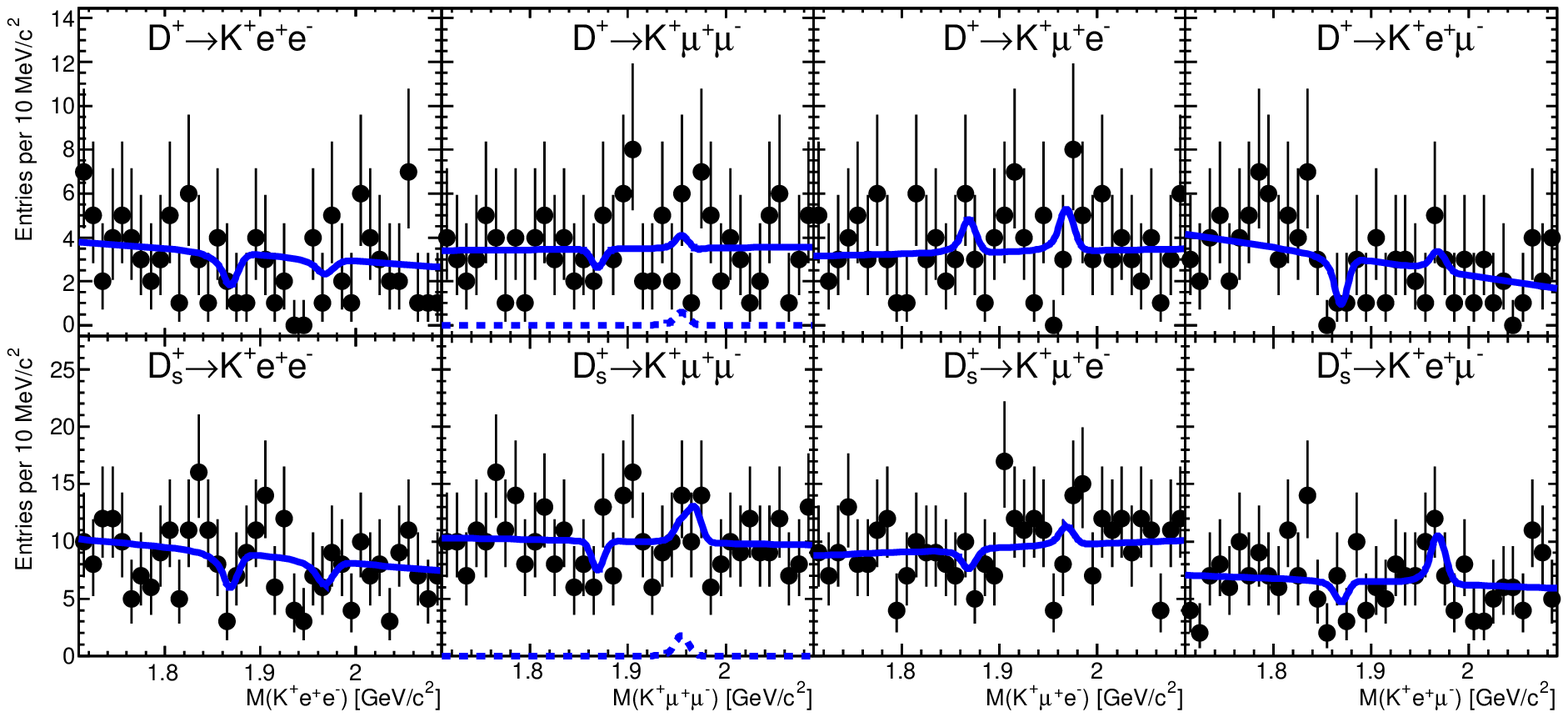}}
\caption{\label{fig:fitDpKll} Invariant-mass distributions for
$\Dp\to\Kp\llpm$ (top) and $\Ds\to\Kp\llpm$ (bottom) candidates.
The solid lines are the results of the fits. 
The background components for the dimuon modes in which muon candidates arise from misidentified hadrons  are shown as dashed curves. }
\end{figure*}

\begin{figure*}[tb]
\centerline{\includegraphics[width=13.5cm]{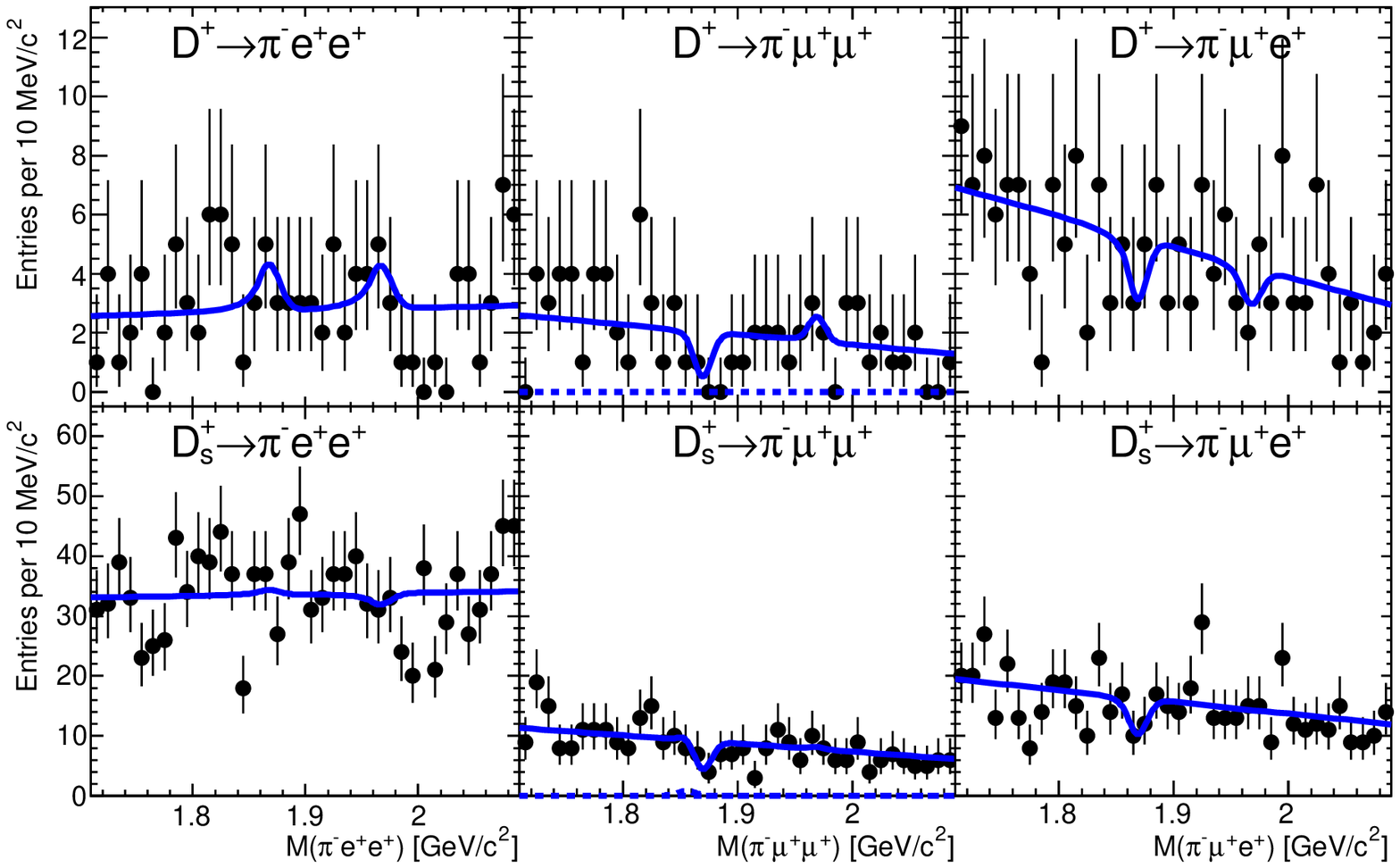}}
\caption{\label{fig:fitDppillLN} Invariant-mass distributions for
$\Dp\to\pim\llpp$ (top) and $\Ds\to\pim\llpp$ (bottom) candidates.
The solid lines are the results of the fits. 
The background component for the dimuon modes in which muon candidates arise from misidentified hadrons  is shown as a dashed curve. }
\end{figure*}

\begin{figure*}[tb]
\centerline{\includegraphics[width=13.5cm]{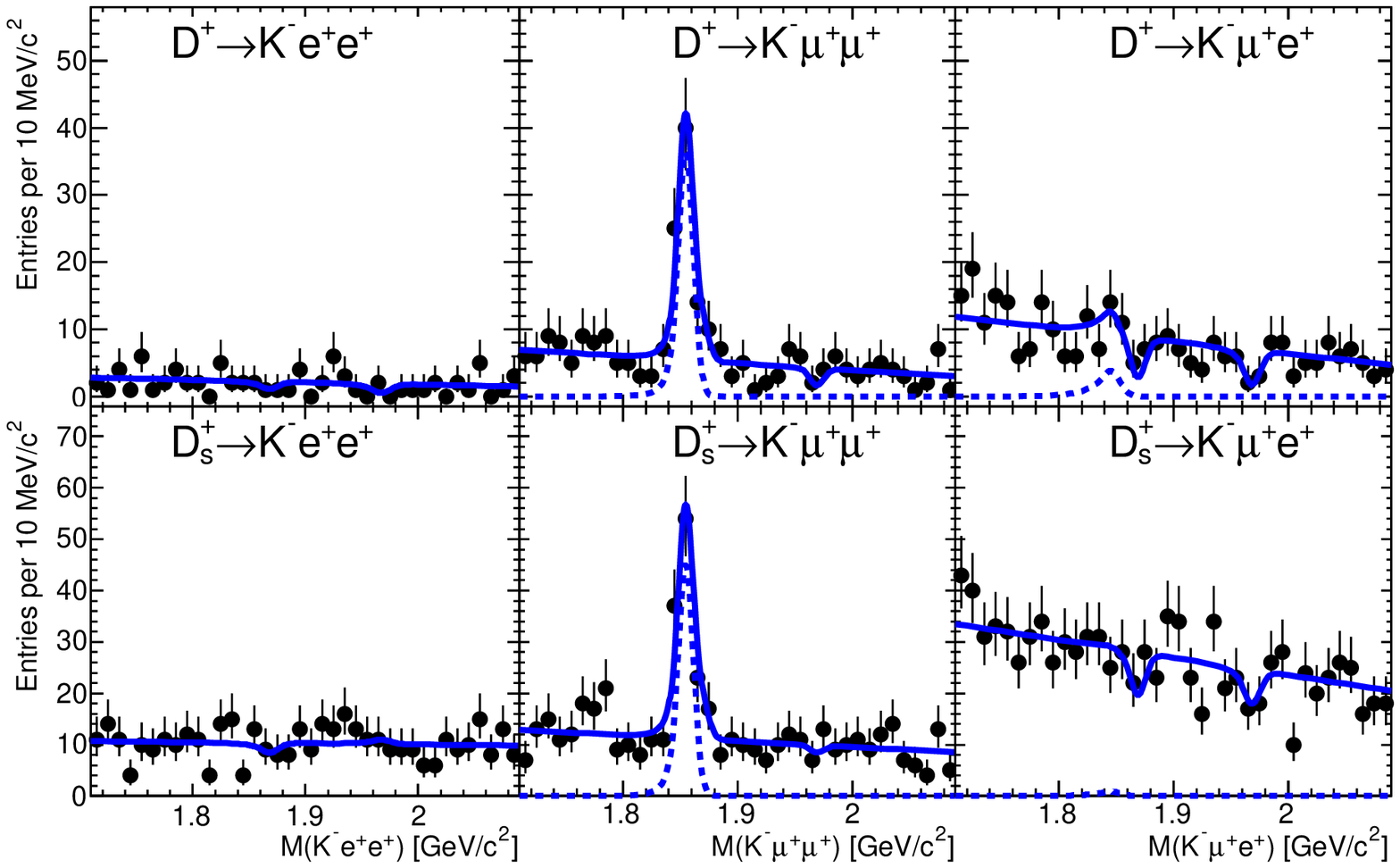}}
\caption{\label{fig:fitDpKllLN} Invariant-mass distributions for
$\Dp\to\Km\llpp$ (top) and $\Ds\to\Km\llpp$ (bottom) candidates.
The solid lines are the results of the fits. 
The background components for the dimuon modes and $\Dsp\to\Km\mup\ep$ in which candidates arise from misidentified hadrons are shown as dashed curves. }
\end{figure*}

\begin{figure*}[tb]
\centerline{\includegraphics[width=18cm]{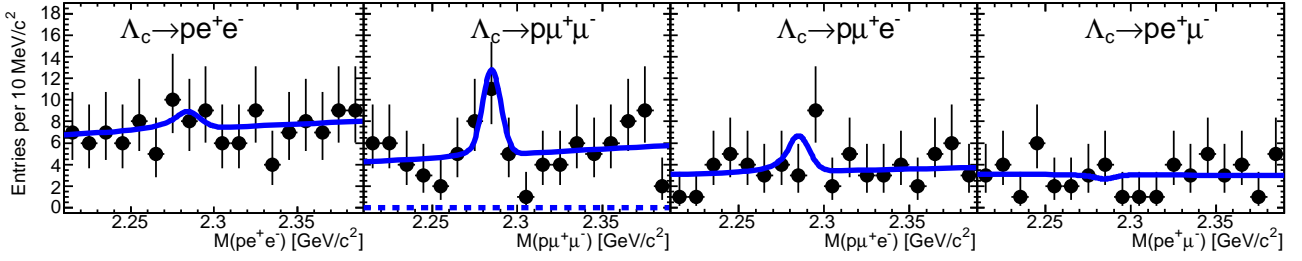}}
\caption{\label{fig:fitLcpll} Invariant-mass distributions for
$\Lc\to\proton\llpm$ candidates.
The solid lines are the results of the fits. 
The background component for the dimuon mode in which muon candidates arise from hadrons misidentified is shown as a dashed curve. }
\end{figure*}

\begin{figure*}[tb]
\centerline{\includegraphics[width=13.5cm]{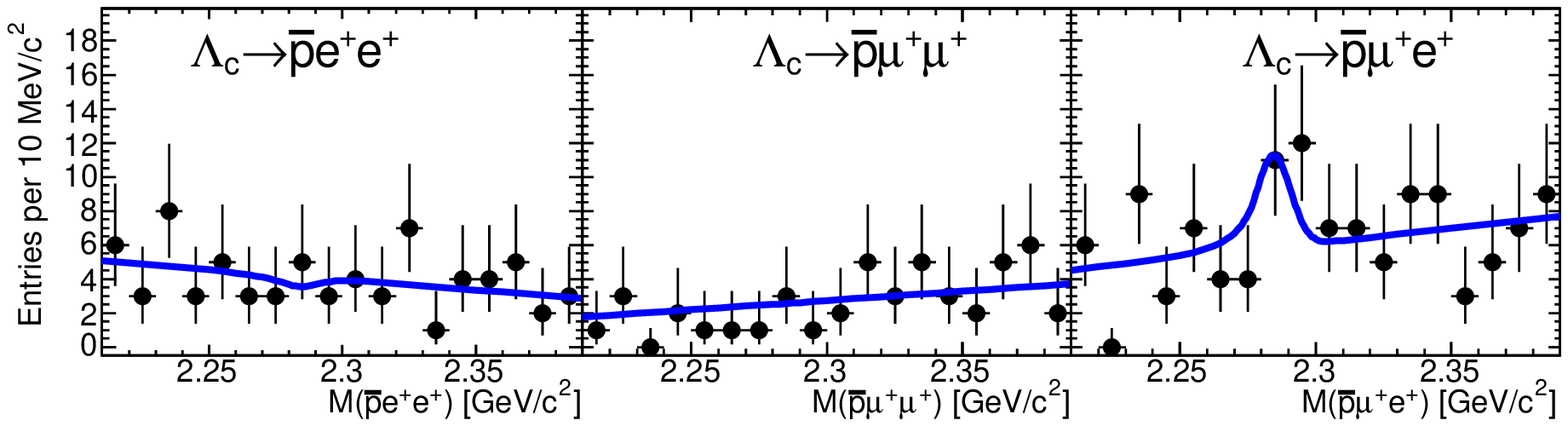}}
\caption{\label{fig:fitLcpllLN} Invariant-mass distributions for
$\Lc\to\antiproton\llpp$ candidates.
The solid lines are the results of the fits. 
}
\end{figure*}

The $h\ell\ell^{(\prime)}$ invariant-mass distributions for signal candidates in all 35 decay
modes are shown in
Figs.~\ref{fig:fitDppill}--\ref{fig:fitLcpllLN}. The signal yields obtained
from the unbinned likelihood fits are listed in Table~\ref{tab:fcncFits}
with statistical and systematic uncertainties. Only systematic
uncertainties associated with the signal and background PDFs are
included in the systematic uncertainty for the yields.  The curves
representing the fits are overlaid in the figures. The most significant signal is seen 
in the distribution for
$\Lc\to\proton\mumu$;
the signal yield has a statistical-only significance of $2.6\sigma$ as determined from 
the change in log-likelihood with respect to zero assumed signal events. With 35 different
measurements, a $2.6\sigma$ deviation is expected with about 25\% probability. 

We calculate upper limits on the branching ratios
\begin{eqnarray*}
&&{{\cal B}(\Dsp\to\pipm\llp)}/{{\cal B}(\Dsp\to\pip\phi)},\\
&&{{\cal B}(\Dsp\to\Kpm\llp)}/{{\cal B}(\Dsp\to\pip\phi)}, \mbox{ and}\\
&&{{\cal B}(\Lc\to\pOrpbar\llp)}/{{\cal B}(\LctopKpi)}
\end{eqnarray*} at 90\% confidence level (CL). 
The upper limits are set using a Bayesian approach with a flat prior for the event yield
in the physical region. The upper limit on the signal yield is defined as the number of 
signal events for which the integral of the likelihood from zero events to that number of events is 90\% of the integral from zero to infinity. The systematic uncertainties are included
in the likelihood as additional nuisance parameters. For
comparison with previous measurements, the upper limits on the
branching fractions at 90\% CL, calculated using
the data of Table~\ref{tab:normBF}, are also given.

For 32 of the 35 decay modes, this analysis is more sensitive
than existing measurements. In most cases, the improvement is
significant (factor of 2 to 60). The largest improvements are seen for
the LFV decays, which are all improved by at least
a factor of 10. The only \Lc decay with a pre-existing limit is
$\Lc\to\proton\mumu$, which we improve by roughly a factor
of 8. For all other \Lc decays this paper presents the first limits.  The
only modes that do not provide a more sensitive limit are
$\Dp\to\pim\ep\ep$, \DtoPimm, and \DstoPimm where existing limits
\cite{Rubin:2010cq,Dzero,focus} are about a factor of two lower than
those presented here.

\begin{figure}[t]
\centerline{\includegraphics[width=9cm]{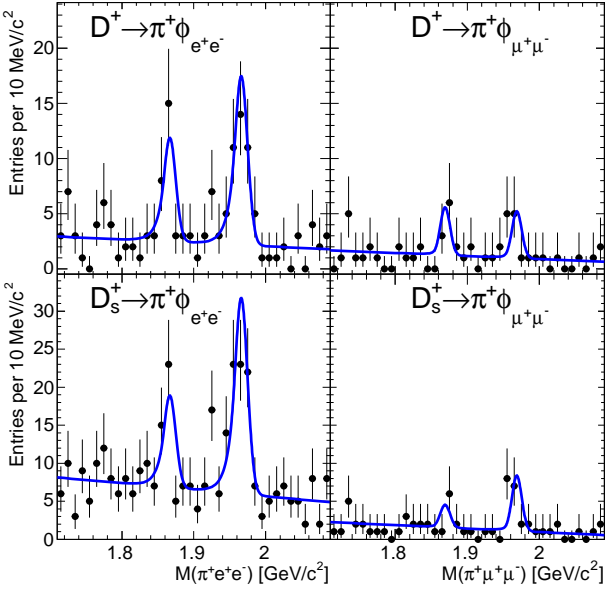}}
\caption{\label{fig:fitDspPiphi} Invariant mass distribution for
(left) $\Dsp\to\pip\phi_{\epem}$ and (right) $\Dsp\to\pip\phi_{\mumu}$
candidates. Only events with dilepton invariant masses in the $\phi$ signal region defined in Sect.~\ref{sec:initialselection} are plotted.
 The solid lines are the results of the
fits. }
\end{figure}

As a cross check, we also examine $\Dsp\to\pip\phi_{\epem}$ and
$\Dsp\to\pip\phi_{\mumu}$ events with dilepton invariant masses in
the $\phi$ signal region defined in
Sect.~\ref{sec:initialselection}. The invariant mass distributions are
shown in Fig.~\ref{fig:fitDspPiphi}. Signals with a statistical
significance of at least $3\sigma$ are seen for all decays except for
$\Dp\to\pip\phi_{\mumu}$. The selection for $\Dp$ and $\Ds$ candidates
differ through the likelihood-ratio criteria, which are optimized
separately; however, signals for both hadrons can be seen with either selection. The mass
distributions are therefore fit allowing for both a $\Dp$ and $\Ds$
signal, but only the signal yield for the hadron for which the
likelihood ratio was constructed is used. Table~\ref{tab:phiModeFits}
gives the fit yields. It also shows the expected yield assuming 
${\cal B}(\phi\to\ellell)={\cal B}(\phi\to\epem)=(2.95\pm0.03)\times
10^{-4}$ \cite{PDG}. The fit yields are in good
agreement with expectations.

\begin{table}
\caption{\label{tab:fcncFits} Signal yields for the fits to the 35
$X_c^+\to h^\pm\llp$ event samples.  The first error is the statistical uncertainty and
the second is the systematic uncertainty. 
The third column lists the estimated signal efficiency.
 The fourth column gives for each signal mode the
90\% CL upper limit (UL) on the ratio of the branching fraction of the signal mode to that of  the normalization mode (BR).
The last column shows the 90\% CL upper limit on the branching fraction for each signal mode (BF). The upper limits include all systematic uncertainties.
}
\begin{center}
\begin{tabular*}{0.48\textwidth}{@{\extracolsep{\fill}}lrrrr} \hline  \hline 
           &          &      & BR UL       & BF UL \\
           & \multicolumn{1}{c}{Yield}    & Eff. & 90\% CL  & 90\% CL \\ 
Decay mode & \multicolumn{1}{c}{(events)} & (\%) & $(10^{-4})$  & $(10^{-6})$ \\ \hline
\rule{0cm}{2.3ex}
$\Dp\to\pip\epem$           & $-3.9\pm1.6\pm1.7$ &   1.56 & 3.9 & 1.1 \\
$\Dp\to\pip\mumu$           & $-0.2\pm2.8\pm0.9$ &   0.46 & 24  & 6.5 \\
$\Dp\to\pip\ep\mun$         & $-2.9\pm3.4\pm2.4$ &   1.21 & 11  & 2.9 \\
$\Dp\to\pip\mup\en$         & $3.6\pm4.3\pm1.3$ &   1.54 & 13  & 3.6 \\
$\Ds\to\pip\epem$           & $8\pm34\pm8$ &   6.36 & 5.4 & 13  \\
$\Ds\to\pip\mumu$           & $20\pm15\pm4$ &   1.21 & 18  & 43  \\
$\Ds\to\pip\ep\mun$         & $-3\pm11\pm3$ &   2.16 & 4.9 & 12  \\
$\Ds\to\pip\mup\en$         & $9.3\pm7.3\pm2.8$ &   1.50 & 8.4 & 20  \\
$\Dp\to\Kp\epem$            & $-3.7\pm2.9\pm3.3$ &   2.88 & 3.7 & 1.0 \\
$\Dp\to\Kp\mumu$            & $-1.3\pm2.8\pm1.1$ &   0.65 & 16  & 4.3 \\
$\Dp\to\Kp\ep\mun$          & $-4.3\pm1.8\pm0.6$ &   1.44 & 4.3 & 1.2 \\
$\Dp\to\Kp\mup\en$          & $3.2\pm3.8\pm1.2$ &   1.74 & 9.9 & 2.8 \\
$\Ds\to\Kp\epem$            & $-5.7\pm5.8\pm2.0$ &   3.20 & 1.6 & 3.7 \\
$\Ds\to\Kp\mumu$            & $4.8\pm5.9\pm1.2$ &   0.85 & 9.1 & 21  \\
$\Ds\to\Kp\ep\mun$          & $9.1\pm6.0\pm2.8$ &   1.74 & 5.7 & 14  \\
$\Ds\to\Kp\mup\en$          & $3.4\pm6.4\pm3.5$ &   2.08 & 4.2 & 9.7 \\
$\Lc\to\proton\epem$        & $4.0\pm6.5\pm2.8$ &   5.52 & 0.8 & 5.5 \\
$\Lc\to\proton\mumu$        & $11.1\pm5.0\pm2.5$ &   0.86 & 6.4 & 44  \\
$\Lc\to\proton\ep\mun$      & $-0.7\pm2.9\pm0.9$ &   1.10 & 1.6 & 9.9 \\
$\Lc\to\proton\mup\en$      & $6.2\pm4.6\pm1.8$ &   1.37 & 2.9 & 19  \\ \hline
$\Dp\to\pim\ep\ep$          & $4.7\pm4.7\pm0.5$ &   3.16 & 6.8 & 1.9 \\
$\Dp\to\pim\mup\mup$        & $-3.1\pm1.2\pm0.5$ &   0.70 & 7.5 & 2.0 \\
$\Dp\to\pim\mup\ep$         & $-5.1\pm4.2\pm2.0$ &   1.72 & 7.4 & 2.0 \\
$\Ds\to\pim\ep\ep$          & $-5.7\pm14.\pm3.4$ &   6.84 & 1.8 & 4.1 \\
$\Ds\to\pim\mup\mup$        & $0.6\pm5.1\pm2.7$ &   1.05 & 6.2 & 14  \\
$\Ds\to\pim\mup\ep$         & $-0.2\pm7.9\pm0.6$ &   2.23 & 3.6 & 8.4 \\
$\Dp\to\Km\ep\ep$           & $-2.8\pm2.4\pm0.2$ &   2.67 & 3.1 & 0.9 \\
$\Dp\to\Km\mup\mup$         & $7.2\pm5.4\pm1.6$ &   0.80 & 37  & 10  \\
$\Dp\to\Km\mup\ep$          & $-11.6\pm4.0\pm3.1$ &   1.52 & 6.8 & 1.9 \\
$\Ds\to\Km\ep\ep$           & $2.3\pm7.9\pm3.3$ &   4.10 & 2.1 & 5.2 \\
$\Ds\to\Km\mup\mup$         & $-2.3\pm5.0\pm2.8$ &   0.98 & 5.3 & 13  \\
$\Ds\to\Km\mup\ep$          & $-14.0\pm8.4\pm2.0$ &   2.26 & 2.4 & 6.1 \\
$\Lc\to\antiproton\ep\ep$   & $-1.5\pm4.2\pm1.5$ &   5.14 & 0.4 & 2.7 \\
$\Lc\to\antiproton\mup\mup$ & $-0.0\pm2.1\pm0.6$ &   0.94 & 1.4 & 9.4 \\
$\Lc\to\antiproton\mup\ep$  & $10.1\pm5.8\pm3.5$ &   2.50 & 2.3 & 16  \\ \hline \hline

\end{tabular*}
\end{center}
\end{table}

\begin{table*}
\caption{\label{tab:phiModeFits} Signal yields for the fits to the
$\Dsp\to\pip\phi_{\ellell}$ candidates.  For the yields, the first uncertainty on the yield is statistical and
the second is systematic. The last column shows the expected number of signal events, where the uncertainty is due to the systematic uncertainty assigned to the efficiency.}
\begin{center}
\begin{tabular*}{0.98\textwidth}{@{\extracolsep{\fill}}lrll} \hline \hline
Decay mode & Yield (events) & Efficiency (\%)  & Expected yield (events) \\ \hline
\rule{0cm}{2.3ex}$\Dp\to\pip\phi_{\epem}$ & $21.8\pm5.8\pm1.5$ & 5.65 & $22.2\pm1.1$ \\
$\Dp\to\pip\phi_{\mumu}$ & $7.5\pm3.4\pm1.4$  & 1.11 & $4.5\pm0.4$ \\
$\Ds\to\pip\phi_{\epem}$ & $63\pm10\pm3$ & 6.46 & $79\pm 3$ \\
$\Ds\to\pip\phi_{\mumu}$ & $12.7\pm4.3\pm2.6$ & 1.07 & $13.1\pm1.2$ \\ \hline \hline

\end{tabular*}
\end{center}
\end{table*}

\section{Conclusions}

\label{sec:conclusions}

Searches for the decay modes $\Dsp\to\pipm\llp$, $\Dsp\to\Kpm\llp$ and
$\Lc\to\pOrpbar\llp$ have been performed using 384\invfb of \epem\
annihilations collected with the \babar\ detector. No signals are observed and we report upper limits on 35 different branching ratios
between
$0.4\times10^{-4}$ and $37\times10^{-4}$ at 90\% CL. This corresponds to limits
on the branching fractions between $1\times 10^{-6}$ and $44\times
10^{-6}$.  These limits are calculated under the assumption of
three-body phase-space decays; the efficiency varies by up to 25\% as
a function of dilepton invariant mass. For 32 of the 35 decay modes studied,
the limits are an improvement over the existing measurements
and therefore provide more stringent constraints on physics beyond the SM.

\section{ACKNOWLEDGMENTS}
\label{sec:Acknowledgments}

We are grateful for the 
extraordinary contributions of our \pep2\ colleagues in
achieving the excellent luminosity and machine conditions
that have made this work possible.
The success of this project also relies critically on the 
expertise and dedication of the computing organizations that 
support \babar.
The collaborating institutions wish to thank 
SLAC for its support and the kind hospitality extended to them. 
This work is supported by the
US Department of Energy
and National Science Foundation, the
Natural Sciences and Engineering Research Council (Canada),
the Commissariat \`a l'Energie Atomique and
Institut National de Physique Nucl\'eaire et de Physique des Particules
(France), the
Bundesministerium f\"ur Bildung und Forschung and
Deutsche Forschungsgemeinschaft
(Germany), the
Istituto Nazionale di Fisica Nucleare (Italy),
the Foundation for Fundamental Research on Matter (The Netherlands),
the Research Council of Norway, the
Ministry of Education and Science of the Russian Federation, 
Ministerio de Ciencia e Innovaci\'on (Spain), and the
Science and Technology Facilities Council (United Kingdom).
Individuals have received support from 
the Marie-Curie IEF program (European Union), the A. P. Sloan Foundation (USA) 
and the Binational Science Foundation (USA-Israel).

\bibliography{paper}  

\end{document}